\documentclass[english,conference,draftcls,onecolumn]{IEEEtran}
\usepackage{draftwatermark}
\SetWatermarkText{DRAFT}
\usepackage[utf8x]{inputenc}
\usepackage{amsmath}
\usepackage{amssymb}
\usepackage[numbers]{natbib}
\usepackage{array}
\usepackage{subfig}
\newcolumntype{P}[1]{>{\centering\arraybackslash}p{#1}}
\makeatletter

\usepackage[english]{babel}
\usepackage{flushend}

\usepackage[colorinlistoftodos]{todonotes}

\begin{document}
\title{Can Network Analysis Techniques help to Predict Design Dependencies? An Initial Study}


\author{
 \IEEEauthorblockN{J. Andr\'{e}s D\'{i}az-Pace\IEEEauthorrefmark{1}, Antonela Tommasel\IEEEauthorrefmark{2}, Daniela Godoy\IEEEauthorrefmark{3}}
 \IEEEauthorblockA{ISISTAN, CONICET-UNICEN. Argentina\\
\IEEEauthorrefmark{1}\emph{andres.diazpace@isistan.unicen.edu.ar},
\IEEEauthorrefmark{2}\emph{antonela.tommasel@isistan.unicen.edu.ar},
\IEEEauthorrefmark{3}\emph{daniela.godoy@isistan.unicen.edu.ar}
}
%
}

\maketitle
\selectlanguage{english}%
\begin{abstract}
The degree of dependencies among the modules of a software system is a key attribute to characterize its design structure and its ability to evolve over time. Several design problems are often correlated with undesired dependencies among modules. Being able to anticipate those problems is important for developers, so they can plan early for maintenance and refactoring efforts. However, existing tools are limited to detecting undesired dependencies once they appeared in the system. In this work, we investigate whether module dependencies can be predicted (before they actually appear). Since the module structure can be regarded as a network, i.e, a dependency graph, we leverage on network features to analyze the dynamics of such a structure. In particular, we apply link prediction techniques for this task. We conducted an evaluation on two Java projects across several versions, using link prediction and machine learning techniques, and assessed their performance for identifying new dependencies from a project version to the next one. The results, although preliminary, show that the link prediction approach is feasible for package dependencies. Also, this work opens opportunities for further development of software-specific strategies for dependency prediction. 
\end{abstract}

\vspace{-0.20cm}
	\section{Introduction}
\label{sec:intro} \vspace{-0.2cm}

As software systems, evolve the amount
and complexity of the interactions among their components is likely to increase, which negatively affects the system design structure \cite{Hochstein:2005:CAD:1709687.1709719}. For instance, certain classes might become coupled because of a new user
feature being added, which makes the corresponding modules dependent on each other.
System degradation symptoms are often related to high coupling and unwanted dependencies, such as: cyclic dependencies, or violations to design rules, among other design smells \cite{conf/qosa/GarciaPEM09}. 
The early detection of such symptoms is important for developers, so that they can plan ahead for actions that preserve the quality of the system.

In this context, there are several tools that help developers to quickly assess if the right dependencies among the system modules are in place, including: LattixDSM, SonarQube, JITTAC or SonarGraph  \cite{Hochstein:2005:CAD:1709687.1709719}. These tools normally extract dependency graphs from the source code and compute different metrics.
Some tools can also bring problems, such as architectural violations or smells, to the developer's attention.
Nonetheless, a limitation of this scenario is that the tools only surface dependencies once they exist in the system. When these problems occur, evidence suggests that developers can be reluctant to fix them \cite{Ali2018}. In a forward-looking scenario, developers would want to know which modules are likely to get coupled in the near future to anticipate dependency-related problems and proactively look for solutions. 

Although there are approaches for computing coupling metrics, very few of them have dealt with the prediction of dependency relations among software components \cite{aryani2014predicting}.  A particular graph-based approach is social networks analysis (SNA), which has been used for modeling both nature and human phenomena. Specifically, SNA techniques can predict links that yet do not exist between pairs of nodes in a network. When it comes to Software Engineering problems, SNA applications
\cite{Bhattacharya:2012:GAP:2337223.2337273} have shown evidence that the topological features of dependency graphs can reveal interesting properties of the software system under analysis.  Nonetheless, link prediction (LP) techniques has not yet been exploited in the Software Engineering community. One exception is \cite{Zhou:2014:BPM:2671850.2671886}, which applied traditional LP techniques for predicting missing dependencies in build configuration files with uneven results. Given the SNA advances over the last years, we argue that LP techniques need to be revisited with respect to (software) dependency graphs.

A first step towards anticipating (unwanted) design dependencies is to assess the predictive performance of LP techniques for general dependencies in different software projects. In this work, we explore the usage of LP for identifying coupling relations between software modules. Our research question (RQ) is \textit{to what extent LP can leverage on information from software versions to predict likely dependencies in the next version}, for those pairs of modules that exist in the analyzed versions. To this end, we report on an initial study with 10 versions of 2 Java projects, which were converted to package dependency graphs.  Although the results with naive LP techniques were not very precise, as expected, we obtained promising results when using LP in tandem with ML models. 

The rest of the article is organized into 4 sections. Section 2 motivates how LP can help in detecting unwanted dependencies. Section 3 describes an experimental study with 2 Java systems, in which we applied 3 approaches for predicting package dependencies. Section 4 discusses the main results. Finally, section 5 covers the conclusions and future work. 

\section{Prediction of Appearing Dependencies}
\label{sec:background}

\begin{figure}
\centering \includegraphics[width=0.80\columnwidth]{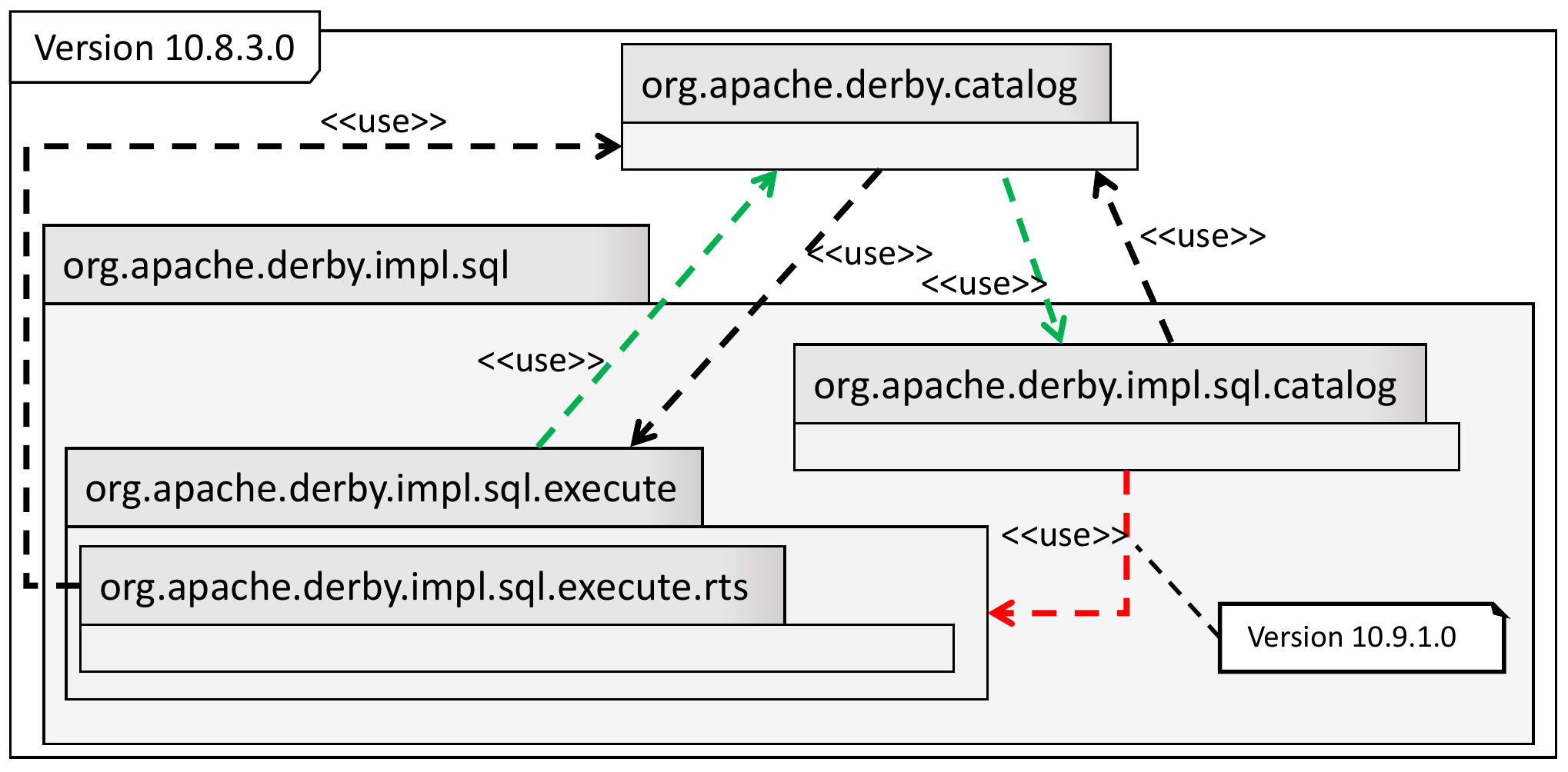}
\caption{Example of Cyclic Dependency (Apache Derby)}\label{fig:example-cyclic-dependency}\vspace{-0.6cm}
\end{figure}

Software systems often exhibit design problems, which can be either introduced during development or along their evolution. A number of these problems manifest themselves as unwanted dependencies in the source code \cite{conf/qosa/GarciaPEM09,Hochstein:2005:CAD:1709687.1709719}. A typical problem is the so-called \textit{Cyclic Dependency}, in which a set of components directly or indirectly depend on each other to function properly. For example, Figure~\ref{fig:example-cyclic-dependency} depicts a cycle (denoted by green and red arrows) among 3 packages of the Apache Derby project (excerpt). The cycle did not exist in version 10.8.3.0 but appeared in version 10.9.1.0 due to a new dependency from \textit{org.apache.derby.impl.sql.catalog} to \textit{org.apache.derby.impl.sql.execute.rts} (red arrow). 

\begin{figure}
\centering \includegraphics[width=0.80\columnwidth]{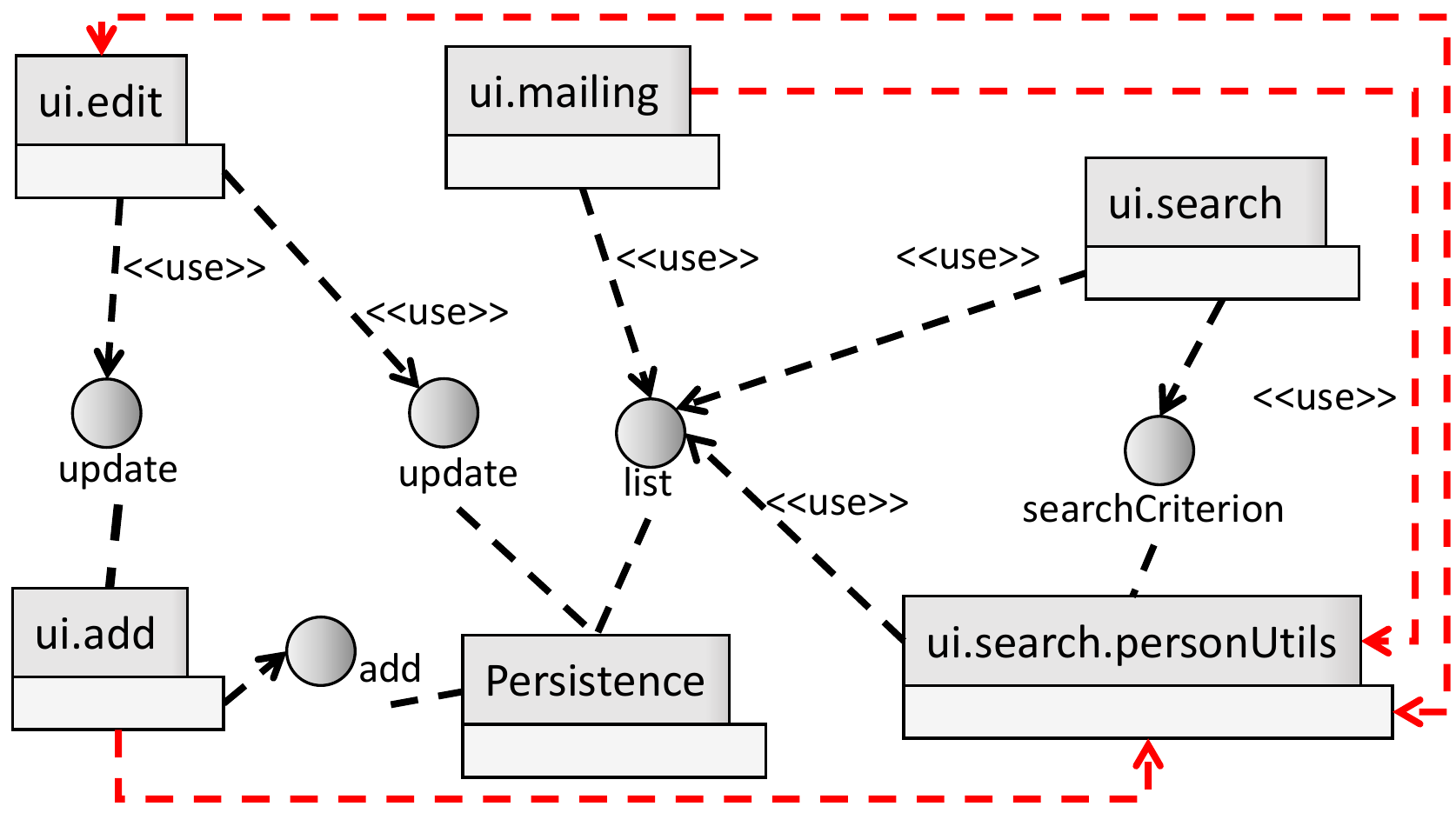}
\caption{Example of Architectural Violation (SubscriberDB)}\label{fig:example-architectural-violation}\vspace{-0.6cm}
\end{figure}

Another common problem is the so-called \textit{Architectural Violation}, which refers to a dependency in the actual (implemented) architecture that was not intended in the original architecture. For example, Figure ~\ref{fig:example-architectural-violation} shows a system called \textit{SubscriberDB}, in which components offer services to other components via interfaces (grey circles), which constitute the allowed architectural interactions. However, 3 code dependencies (red arrows) violated those interaction rules. 

For the Java examples above, we assume an architecture compliance process that periodically checks whether the current system implementation satisfies a set of design rules, and reports any issues to developers.  Dependencies judged as "undesirable" might come from new functionality in existing classes or new classes being created under existing packages. These dependencies get inadvertently introduced in the code. Furthermore, despite class-level changes, the package structure remains more or less stable across system versions, while package dependencies keep being added. There are a few exceptions, such as: the initial versions, in which the main structure and functionality are fleshed out, or a version with a major refactoring. Overall, looking at the package network and its evolution, it is possible (and also beneficial) to predict dependencies between packages that are yet unconnected. 

In this context, we resort to \textit{link prediction} (LP) techniques, which adapts SNA for studying the evolution of a network using models of network \textit{features} \cite{Liben-Nowell:2007:LPS:1241540.1241551}. LP seeks to infer "missing" links between pairs of nodes  based on their observable links and attributes. 
A prerequisite for applying LP is to transform the system under analysis
into a dependency graph. More formally, a dependency graph is a graph
$DG\left(V,E\right)$, where each node $v\in V$ is a module, and
each edge (or link) $e\left(v,v'\right)\in E$ is a dependency from
nodes $v$ to $v'$ ($v,v'\in V$). Since we deal with Java systems,
nodes correspond to packages while edges represent usage relations
between packages. 
Our LP task takes a $DG\left(V,E\right)$ at time $n$, and then infers the edges that will be added to $DG\left(V,E\right)$ in time $n+1$. More formally, let $U$ be the set of all possible edges among nodes in $DG\left(V,E\right)$. The LP task generates a list $R$ of all possible edges in $U - E$, and indicates whether each edge is present in $DG\left(V,E\right)$ at $n+1$.

LP in social networks is based on the principle of homophily~\cite{doi:10.1146/annurev.soc.27.1.415}, which states that interactions between similar individuals occur at a higher rate than those among dissimilar ones. In our context, this means that similar packages (according to some criteria) have a higher chance to establish dependencies than dissimilar packages. Most techniques for the LP problem  use graph topological features that assess similarity between pairs of nodes \cite{Liben-Nowell:2007:LPS:1241540.1241551}. These metrics normally produce a  ranking of edges in $R$. 
Table~\ref{tab:Cycles-prediction} shows rankings of the top-5 predicted dependencies for \textit{org.apache.derby.impl.sql.catalog} (example of Figure~\ref{fig:example-cyclic-dependency}), according to different similarity metrics. Note that \textit{org.apache.derby.impl.sql.execute.rts} is included in both rankings (i.e., the prediction is positive) but in different positions.  Although intuitive, this ranking-based approach is not always effective for complex networks.  A more comprehensive approach is to cast LP to a \textit{classification problem} in which network features are used to build a prediction model. 

\begin{table}
\caption{\label{tab:Cycles-prediction} Examples of top-5 dependency rankings for a package}
\begin{centering}
\begin{tabular}{|c|c|c|}
\hline 
 & \emph{\footnotesize Common Neighbours} & \emph{\footnotesize Adamic Adar}\tabularnewline
\hline 
{\scriptsize 1} & {\scriptsize org.apache.derby.impl.sql} & \textbf{\scriptsize org.apache.derby.impl.sql.execute}\tabularnewline
\hline 
{\scriptsize 2} & \textbf{\scriptsize org.apache.derby.impl.sql.execute} & {\scriptsize org.apache.derby.impl.sql}\tabularnewline
\hline 
{\scriptsize 3} & {\scriptsize org.apache.derby.impl.sql.conn} & {\scriptsize org.apache.derby.impl.sql.conn}\tabularnewline
\hline 
{\scriptsize 4} & {\scriptsize org.apache.derby.impl.db} & {\scriptsize org.apache.derby.impl.db}\tabularnewline
\hline 
{\scriptsize 5} & {\scriptsize org.apache.derby.impl.store.raw.data} & {\scriptsize org.apache.derby.impl.jdbc}\tabularnewline
\hline 
\end{tabular}
\par\end{centering}

\vspace{-0.6cm}

\end{table}
\vspace{-0.07cm}
\section{Study Settings}

In order to assess the performance of LP techniques, we took a list of 10 versions for 2 Java systems (see Table~\ref{tab:versions}), and tried to predict package dependencies using different approaches. The systems, \textit{SubscriberDB} (SDB) ($\sim$10 KLOC)  
and \textit{HealthWatcher} (HW) \cite{DBLP:conf/sbcars/VidalGOGPM16} ($\sim$49 KLOC), 
 were chosen because we had first-hand knowledge about their evolution and version issues. For this work, we analyzed package dependencies in general, rather than dependencies related to specific design problems. The dependency graphs for the versions were computed with the CDA tool\footnote{http://www.dependency-analyzer.org}. Dependencies among classes were ignored. For each version $v$, we looked into the number of package dependencies (sparsity) and the amount of code changes from $v$ to the next version. For LP to produce reasonable outputs, a pair of consecutive versions ($v_n$, $v_{n+1}$) should meet some conditions: i) both $v_n$ and $v_{n+1}$ have almost the same number of packages, ii) $v_{n+1}$ has changes in their classes or adds new classes, and iii) a percentage of dependencies is added in $v_{n+1}$. These filtering conditions yielded a subset of the versions (in bold in Table~\ref{tab:versions}). We should also note that potentially appearing dependencies for new packages added in $v_{n+1}$ were disregarded.  
\begin{table}
\caption{\label{tab:versions}Summary of versions for SDB and HW}
\begin{centering}
\setlength{\tabcolsep}{2pt}%
\begin{tabular}{|c|c|c|c|c|c|c|c|c|c|c|}
\cline{1-5} \cline{7-11} 
 & \emph{\scriptsize \#c} & \emph{\scriptsize \#p} & \emph{\scriptsize \#deps} & \emph{\scriptsize sparsity} &  &  & \emph{\scriptsize \#c} & \emph{\scriptsize \#p} & \emph{\scriptsize \#deps} & \emph{\scriptsize sparsity}\tabularnewline
\cline{1-5} \cline{7-11} 
\emph{\scriptsize HWv1} & {\scriptsize 88} & {\scriptsize 19} & {\scriptsize 67} & {\scriptsize 0.8041} &  & \emph{\scriptsize SDBv1} & {\scriptsize 98} & {\scriptsize 14 } & {\scriptsize 30} & {\scriptsize 0.8352}\tabularnewline
\cline{1-5} \cline{7-11} 
\textbf{\emph{\scriptsize HWv2}} & \textbf{\scriptsize 92} & \textbf{\scriptsize 20} & \textbf{\scriptsize 70 (+8, -5)} & \textbf{\scriptsize 0.8157} &  & \textbf{\emph{\scriptsize SDBv2}} & \textbf{\scriptsize 167} & \textbf{\scriptsize 16} & \textbf{\scriptsize 47 (+17)} & \textbf{\scriptsize 0.8042}\tabularnewline
\cline{1-5} \cline{7-11} 
\textbf{\emph{\scriptsize HWv3}} & \textbf{\scriptsize 104} & \textbf{\scriptsize 21} & \textbf{\scriptsize 75 (+5)} & \textbf{\scriptsize 0.8214} &  & \textbf{\emph{\scriptsize SDBv3}} & \textbf{\scriptsize 192} & \textbf{\scriptsize 17} & \textbf{\scriptsize 50 (+4, -1)} & \textbf{\scriptsize 0.8162}\tabularnewline
\cline{1-5} \cline{7-11} 
\textbf{\emph{\scriptsize HWv4}} & \textbf{\scriptsize 106} & \textbf{\scriptsize 22} & \textbf{\scriptsize 85 (+10)} & \textbf{\scriptsize 0.8160} &  & \emph{\scriptsize SDBv4} & {\scriptsize 193} & {\scriptsize 17} & {\scriptsize 50} & {\scriptsize 0.8162}\tabularnewline
\cline{1-5} \cline{7-11} 
\textbf{\emph{\scriptsize HWv5}} & \textbf{\scriptsize 108} & \textbf{\scriptsize 22} & \textbf{\scriptsize 86 (+7, -2)} & \textbf{\scriptsize 0.8138} &  & \emph{\scriptsize SDBv5} & {\scriptsize 193} & {\scriptsize 17} & {\scriptsize 50} & {\scriptsize 0.8162}\tabularnewline
\cline{1-5} \cline{7-11} 
\emph{\scriptsize HWv6} & {\scriptsize 112} & {\scriptsize 23} & {\scriptsize 91} & {\scriptsize 0.8201} &  & \emph{\scriptsize SDBv6} & {\scriptsize 193} & {\scriptsize 17} & {\scriptsize 50} & {\scriptsize 0.8162}\tabularnewline
\cline{1-5} \cline{7-11} 
\emph{\scriptsize HWv7} & {\scriptsize 116} & {\scriptsize 23} & {\scriptsize 91} & {\scriptsize 0.8201} &  & \emph{\scriptsize SDBv7} & {\scriptsize 195} & {\scriptsize 17} & {\scriptsize 50} & {\scriptsize 0.8162}\tabularnewline
\cline{1-5} \cline{7-11} 
\textbf{\emph{\scriptsize HWv8}} & \textbf{\scriptsize 120} & \textbf{\scriptsize 24} & \textbf{\scriptsize 96 (+5)} & \textbf{\scriptsize 0.8261} &  & \textbf{\emph{\scriptsize SDBv8}} & \textbf{\scriptsize 195} & \textbf{\scriptsize 17} & \textbf{\scriptsize 51 (+1)} & \textbf{\scriptsize 0.8125}\tabularnewline
\cline{1-5} \cline{7-11} 
\textbf{\emph{\scriptsize HWv9}} & \textbf{\scriptsize 132} & \textbf{\scriptsize 24} & \textbf{\scriptsize 97 (+1)} & \textbf{\scriptsize 0.8242} &  & \emph{\scriptsize SDBv9} & {\scriptsize 195} & {\scriptsize 17} & {\scriptsize 51} & {\scriptsize 0.8125}\tabularnewline
\cline{1-5} \cline{7-11} 
\textbf{\emph{\scriptsize HW 10}} & \textbf{\scriptsize 135} & \textbf{\scriptsize 25} & \textbf{\scriptsize 101 (+4)} & \textbf{\scriptsize 0.8317} &  & \emph{\scriptsize SDBv10} & {\scriptsize 195} & {\scriptsize 17} & {\scriptsize 51} & {\scriptsize 0.8125}\tabularnewline
\hline 
\multicolumn{11}{>{\centering}p{0.55\columnwidth}}{{\scriptsize where "\#c" indicates number
of classes, "\#p" number of packages,
"\#deps" number of dependencies, "+"
indicates the number of added dependencies, and "-"
indicates the number of disappeared dependencies }}\tabularnewline
\end{tabular}
\par\end{centering}

\vspace{-0.7cm}
\end{table}

Dependencies were predicted based on 3 approaches with increased complexity. 
These approaches relied on the same topological metrics, which provide global or local information of the network~\cite{Liben-Nowell:2007:LPS:1241540.1241551}. The metrics considered were: \emph{Adamic-Adar}, \emph{Common Neighbours}, \emph{Katz Score}, \emph{Resource Allocation}, \emph{SimRank}, and \emph{S{\o}rensen}. The best performing metrics from ~\cite{seke2013b}: \emph{Kulczynski}, \emph{RelativeMatching}, and \emph{RusselRao}, were also added. Next, we describe each approach in detail. 

\subsubsection{Ranking-based LP} 

This approach follows directly from the homophily principle, and gives a baseline for the study. For a package $p$, a ranking of packages is built, based on their chance of having a future dependency with $p$, according to a similarity metric. For pairs of consecutive versions, the quality of predictions was evaluated in terms of precision (i.e., the ratio of actual dependencies discovered to the total number of predictions) for the top-$N$ dependencies of the ranking.


\subsubsection{Training a Classification Model}

In this ML approach, a binary classifier is trained using the topological information provided by a given graph version. An instance for the classifier consists of: a pair of nodes, a list of features (e.g., structural metrics) for the pair, and a label indicating if the nodes are linked (positive class) or not (negative class). Existing dependencies are used to compute features for instances of the positive class, while missing dependencies are used to compute features for instances of the negative class.
Both training and test sets need to be defined. The training set considers the full graph for $v_n$, and the test set considers the full
graph (i.e. the real distribution of links) for $v_{n+1}$.
 The prediction (i.e., classification) of dependencies was made with the Weka implementation of SMO 
\footnote{https://www.cs.waikato.ac.nz/ml/weka/}
parameterized with a RBF kernel, which is useful for unbalanced instance sets as it is the case of the LP problem. 
Since traditional classification metrics (e.g., precision, recall) might not be sufficient for correctly analyzing this scenario~\cite{Yang:2015:ELP:2837335.2837363}, 
performance was assessed by means of the area under the
precision-recall curve (AUPR). 

\subsubsection{Time Series Forecasting}

This approach combines dynamic
SNA (i.e., observations of the graph at different time periods) with topological features to learn a robust ML model able to predict new links~\cite{Rossetti2016}. It also relies on communities as closely-related nodes of the graph.
Prediction is based on a classifier trained with the last known version of the system, $v_n$. The test set considers the estimated feature scores for $v_{n+1}$. The forecasting algorithm was implemented in Weka. Like in the previous approach, we evaluated performance with AUPR.\vspace{-0.1cm}

\section{Evaluation of Predictive Performance} 
\label{sec:evaluation-results} 



The precision results of the ranking-based approach in both projects were in the range 0.14-0.25 at most, thus predictions were not satisfactory. These results are in line with those in \cite{Zhou:2014:BPM:2671850.2671886}. We believe this is due to relying only on the homophily principle, which does not always hold for software modules. For instance, two similar packages can intentionally be designed to not become dependent on each other, based on business logic or
modularity reasons. 
On the contrary, dependencies
might still appear between dissimilar packages. This evidence motivates approaches able to learn "exceptions" to homophily.
\begin{figure}
\includegraphics[width=0.95\columnwidth]{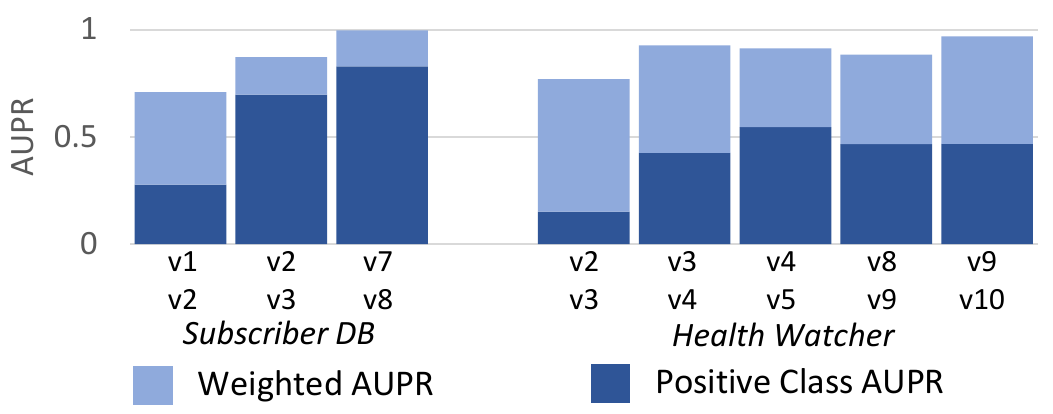}
\caption{AUPR for selected versions using a binary classifier}\label{fig:supervised} \vspace{-0.55cm}
\end{figure}
\begin{figure*}
\begin{centering}
\subfloat[Health Watcher]{\centering{}\includegraphics[width=0.95\textwidth]{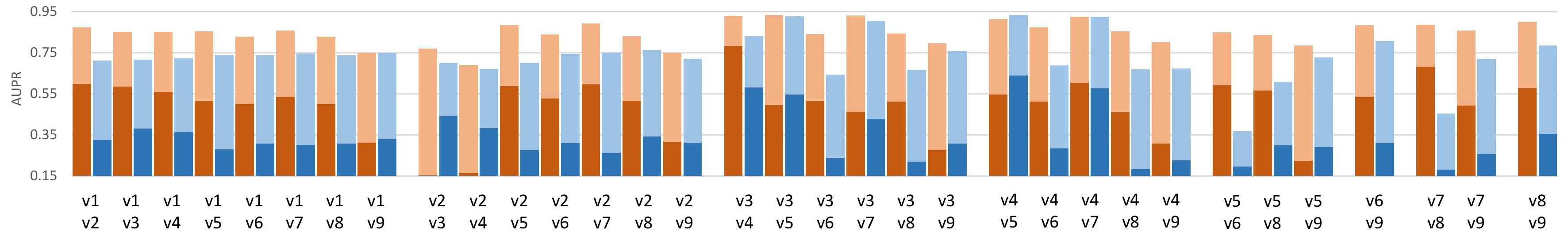}}\vspace{-0.4cm}
\par\end{centering}
\begin{centering}
\subfloat[Subscriber DB]{\begin{centering}
\includegraphics[width=0.95\textwidth]{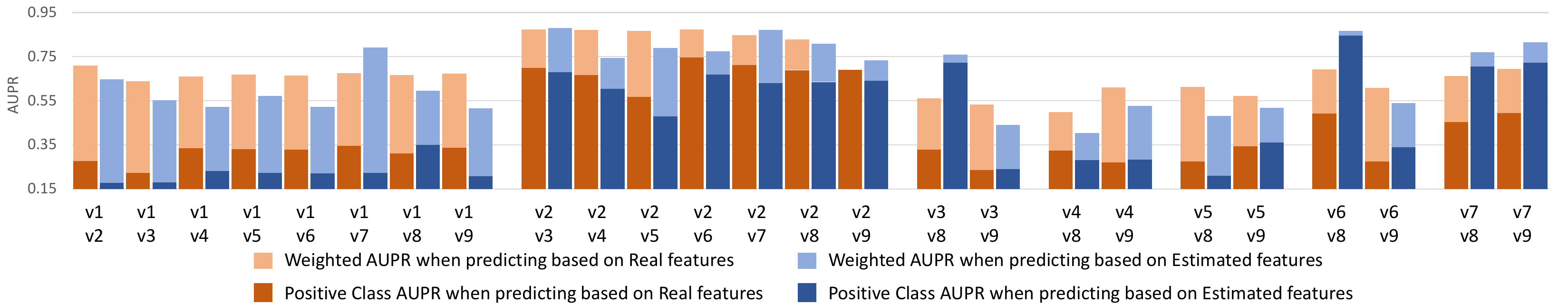}
\par\end{centering}
}\vspace{-0.2cm}
\par\end{centering}
\caption{\label{fig:Forecast}AUPR for selected version pairs using time series forecasting}\vspace{-0.8cm}
\end{figure*}
In the second approach, 
we computed AUPR for the positive class (i.e., existing dependencies) and also a weighted AUPR considering both classes (i.e., existing and non-existing dependencies). The values for a binary classifier over selected versions of SDB and HW are depicted in Figure~\ref{fig:supervised}. For each pair (X axis), the first item is the version for the training set, while the second item is the version for the test set. Only pairs with new dependencies between existing packages are shown. We obtained better precision for the weighted class, with average values of 0.85 and 0.96 for SDB and HW. Nonetheless, precision for the positive class 
was still far from ideal, with values of 0.74 and 0.23 respectively.




Despite a high weighted AUPR, Figure~\ref{fig:supervised} shows variations in the positive class AUPR. For instance, the low AUPR for the positive class in $v_2$-$v_3$ for HW means that the classifier finds all new dependencies correctly (good recall), but it also mistakenly reports non-existing dependencies (low precision due to false positives). Thus, in practice, the set of predicted dependencies might be noisy and very long to analyze by developers. Nonetheless, for the SDB pairs $v_2$-$v_3$ and $v_7$-$v_8$, the trained model achieved both good precision and recall. 
The performance variations imply that, for certain versions, it is difficult to differentiate between dependencies and non-dependencies due to similar structural characteristics. For example, the positive class AUPR consistently increased as HW evolved. Similar trends were noticed for SDB, although with higher values than for HW. For SDB, only new dependencies were added in $v_2$-$v_3$, while in $v_1$-$v_2$ existing dependencies disappeared. These facts reinforce our position about the need to consider additional information for having good predictions.


To improve the positive class AUPR, we exercised the third approach based on forecasting. Figure~\ref{fig:Forecast} presents the AUPR results for selected versions of SDB and HW. For each pair (X axis), the versions represent the span for the estimations, e.g., $v_1$-$v_3$ means that $v_1$, $v_2$ and $v_3$ served to estimate the features for $v_4$, which was the test set. The results compare predictions with the real (in red) and estimated (in blue) features. As observed, real features still expose difficulties for identifying dependencies and non-dependencies, which leads to performance variability
across the versions. 
For the estimated features, in turn, we see
the same or even better predictions than for the real features,
allowing the classifier to achieve good performance. In fact, the average precision for the positive class increased regarding the second approach to 0.84 and 0.37, for SDB and HW respectively.  
The choice of versions for forecasting seems relevant. When starting the estimation in $v_1$ for SDB, results are lower than when starting in $v_2$. Similarly
for HW, results are lower when starting in $v_2$ than in $v_3$. Moreover, as the number of HW versions increased, the quality of predictions decreased as for $v_2$-$v_4$ and
$v_2$-$v_9$. This effect can be related to the information or the structural changes in each version, regardless of the actual number of versions.
For instance, if a version undergoes a refactoring that greatly affects dependencies, information from prior versions might not be representative of the actual system structure. 

\vspace{-0.1cm}
\section{Conclusions and Outlook}
\label{sec:conclusions} 

In this work, we make the case that usage dependencies among software modules can be predicted by leveraging on LP and topological information from system versions.
Our goal was not to develop the "best predictor", but rather to assess i) the LP performance in dependency graphs, and ii) the kind of information required for having reasonable predictions. Although naive LP techniques are not adequate for the task, we obtained evidence that combining them with ML techniques improves their performance. This approach is interesting for dependency management and architecture compliance tools, as it helps to anticipate dependency-related design problems. 

Despite the potential of LP techniques, further investigation is needed. A systematic study with more systems is required to corroborate our initial findings. The features currently used in the approaches can be extended. In particular, software-specific metrics and similarity criteria (e.g., for source code artifacts) are necessary. Also, the concept of communities~\cite{Rossetti2016} (used in time series forecasting) can help boosting predictions. 
In addition, we envision the development of a tool able to infer unwanted dependencies among software elements. Customized LP algorithms for dependency-based design problems (e.g., layering violations, cycles, or hub-like dependencies) can be created. 
Design metrics can be also estimated (e.g., instability, or change proneness). At last, integration with existing tools, such as SonarQube,  is another subject for future work.

\vspace{-0.2cm}


\footnotesize
\bibliographystyle{plainnat}

\end{document}